\documentclass[epsfig, 12pt, onecolumn]{article}
\usepackage[sumlimits]{amsmath}
\usepackage{amsfonts,amssymb,latexsym,eufrak,mathrsfs,graphicx,enumerate,bbm}
\renewcommand{\i}{\ensuremath{\text{i}}}
\renewcommand{\d}{\ensuremath{\text{d}}}

\newcommand{\e}{\ensuremath{\text{e}}}
\newcommand{\D}{\ensuremath{\text{D}}}
\newcommand{\Dbar}{\ensuremath{\bar{\text{D}}}}
\newcommand{\J}{\ensuremath{\text{J}}}
\newcommand{\Jd}{\ensuremath{{\text{J}}^{\dagger}}}

\newcommand{\Phid}{\ensuremath{\Phi^{\dagger}}}
\newcommand{\Lambdad}{\ensuremath{\Lambda^{\dagger}}}
\newcommand{\Hd}{\ensuremath{H^{\dagger}}}

\newcommand{\thetab}{\bar{\theta}}
\newcommand{\alphad}{\dot{\alpha}}
\newcommand{\betad}{\dot{\beta}}

\newcommand{\newc}{\newcommand}
\def\beq{\begin{equation}}
\def\eeq{\end{equation}}
\def\bea{\begin{eqnarray}}
\def\eea{\end{eqnarray}}
\newc{\ie}{{\it i.e.}}          \newc{\etal}{{\it et al. }}
\newc{\eg}{{\it e.g.}}          \newc{\etc}{{\it etc.}}
\newc{\cf}{{\it c.f.}}

\begin{document}
\thispagestyle{empty}
\vspace*{.5cm}
\noindent
\hspace*{\fill}{\large OUTP-04/25P}\\
\vspace*{2.0cm}

\begin{center}
{\Large\bf Pseudo-Goldstones from Supersymmetric Wilson Lines on 5D Orbifolds}
\\[2.5cm]
{\large Thomas Flacke, Babiker Hassanain, and John March-Russell
}\\[.5cm]
{\it Rudolf Peierls Centre for Theoretical Physics,\\
University of Oxford, 1 Keble Road, Oxford OX1 3NP, UK}
\\[.2cm]
(March 2005)
\\[1.1cm]

{\bf Abstract}
\end{center}
We consider a U(1) gauge theory on the five dimensional orbifold
$\mathcal{M}_4\times S^1/Z_2$, where $A_5$ has even $Z_2$ parity. This leads to a light pseudoscalar degree of freedom $W(x)$
in the effective 4D theory below the compactification scale
arising from a gauge-invariant brane-to-brane Wilson line.
As noted by Arkani-Hamed \etal in the non-supersymmetric $S^1$ case the 5D bulk
gauge-invariance of the underlying theory together with the non-local
nature of the Wilson line field leads to the protection of the
4D theory of $W(x)$ from possible large global-symmetry
violating quantum gravitational effects.  We study the $S^1/Z_2$
theory in detail, in particular developing the supersymmetric generalization
of this construction, involving a pseudoscalar Goldstone field (the `axion')
and its scalar and fermion superpartners (`saxion' and `axino'). The
global nature of $W(x)$ implies the absence of independent
Kaluza-Klein excitations of its component fields.  The
non-derivative interactions of the (supersymmetric) Wilson line in
the effective 4D theory arising from U(1) charged 5D fields $\Phi$ propagating between the boundary branes are studied.
We show that, similar to the non-supersymmetric $S^1$ case, these interactions are suppressed by $\exp(-\pi R m_{\Phi})$ where $\pi R$ is the size of the extra dimension.

\noindent

\newpage

\setcounter{page}{1}

\section{Introduction}

The study of both non-supersymmetric and supersymmetric gauge field theories on
the five dimensional manifold $\mathcal{M}_4\times S^1$ and orbifolds
$\mathcal{M}_4\times S^1/Z_2$ and $\mathcal{M}_4\times
S^1/(Z_2\times Z'_2)$ has attracted much recent interest. 
One application of gauge theories on $\mathcal{M}_4\times S^1/(Z_2\times Z'_2)$
has been to provide novel solutions to the problems of standard 4D supersymmetric
grand-unified (GUT) theories~\cite{orbguts,orbgutsunif,su5,so10}. The
GUT gauge symmetry is now realized in 5 or more space-time 
dimensions and broken to the Standard Model (SM) gauge group by
compactification on an orbifold, 
utilizing boundary conditions that violate the GUT-symmetry. In the most 
studied case of 5 dimensions both the GUT group and 5D supersymmetry are 
broken by compactification on $S^1/(Z_2\times Z_2')$, leading to a 4D
N=1 SUSY model with SM gauge group.  This construction provides elegant
solutions to the problems of conventional GUTs with Higgs breaking, including 
doublet-triplet splitting, dimension-5 proton decay, and Yukawa 
unification in the first two generations, while maintaining, at least at 
leading order, the desired gauge coupling unification~\cite{orbgutsunif,CPRT}.
The hierarchy between the strong coupling scale $M$ of the 5D gauge 
theory and the compactification scale $1/R$ $(MR\sim 10^2\cdots 10^3)$ can
also be used to generate a fermion mass hierarchy~\cite{HMROS}. 

In another line of development, Arkani-Hamed \etal~\cite{AHCCR}
studied a 5D $U(1)$ gauge theory compactified on $\mathcal{M}_4\times S^1$
and the gauge-invariant Wilson line phase $\Theta$
associated to the zero mode of $A_5$
\begin{equation}
\exp(-\i\Theta) \equiv \exp\left( -\i\int \d y A_5 \right) .
\end{equation}
The phase $\Theta$ becomes a field in the 4D effective theory 
when the $A_5$ zero-mode is allowed to have a slow variation as a function of the
coordinates on $\mathcal{M}_4$.  The associated 4D field theory has the
interesting property that it inherits a global shift symmetry
$\Theta\rightarrow \Theta + c$ from the 5D underlying gauge symmetry, 
and this leads to the phenomenologically interesting possibility that the
global shift symmetry is protected from dangerous local 5D quantum-gravitational
symmetry violating effects \cite{quantgrav} by the underlying 5D gauge symmetry. In particular,
Arkani-Hamed \etal~\cite{AHCCR} utilized this construction to build interesting
new models of inflation that at least in part address the problems of traditional
4D inflationary model building.  Later papers \cite{Kaplan,Hofmann}
generalized this to supersymmetric models of inflation on
$\mathcal{M}_4\times S^1$ and to applications to quintessence
and axion model building~\cite{Pilo,Choi}.  In our present work we wish to further generalize
these ideas to the supersymmetric case on $Z_2$ orbifolds of $\mathcal{M}_4 \times S^1$,
bringing the models closer to the orbifold GUT constructions.  

In the orbifold case the most commonly studied situation is where the orbifold parity
operation is chosen such that the $A_5$ component of the 5D gauge field is odd, implying
the absence of a $A_5$ zero mode (the $A_\mu$ component is then even, thus
leading to an $A_\mu$ zero mode and thus a 4D gauge theory). 
However as argued in Ref.\cite{Choi} in the context of 5D theories of
axions the other choice, where the 4D gauge zero mode $A_\mu$ is projected
out by taking $A_5$ to be \emph{even} is also phenomenologically interesting.  
When $A_5$ has even $Z_2$ parity, there exists a zero
mode $A_5(x,y)=A_5(x,0)$ in the spectrum of the theory.
In this paper, we consider a $U(1)$ gauge field on the five
dimensional orbifold $\mathcal{M}_4\times S^1/Z_2$, where we
choose $A_5$ to have even $Z_2$ parity, leading to a light
physical gauge-invariant Wilson line degree of freedom in the
effective 4D theory below the compactification scale.\footnote{Our treatment
 trivially extends to a compactification on $S^1/(Z_2\times Z'_2)$ with $A_5$ even at both fixed points.}  Similar
to the $S^1$ case \cite{AHCCR} the 5D bulk
gauge-invariance of the underlying theory together with the non-local
nature of the Wilson line field leads to the protection of the
4D theory of the Wilson line field from possible global-symmetry
violating quantum gravitational effects, at least if the size of the
5th dimension, $\pi R$ is large compared to the fundamental 5D Planck scale
$M_5$.  In the following sections, we study the $S^1/Z_2$ theory in detail,
in particular developing the supersymmetric generalization
of this construction, and we explore the effect of the Wilson line on the
effective 4D theory in analogy to the case of $S^1$ \cite{AHCCR}.

\section{Non-supersymmetric Wilson lines on \boldmath$S_1 / Z_2$}\label{flat}

We want to study Wilson lines arising from the fifth component, $A_5$, of
a $U(1)$ 5D gauge field $A_M$ and its coupling to bulk bosons and
fermions $\Phi$, $\Psi$ and brane bosons and fermions $\phi$, $\psi$.
(Our conventions are given in Appendix A.)
In order to have a zero-mode, the $A_5$ component
must have even $Z_2$ parity $A_5(-y) = A_5(y)$
which fixes the parities of the gauge
parameter $\Lambda$ and the $A_{\mu}$ components to be odd due to
the gauge transformation $A_M\rightarrow A_M + \partial_M\Lambda$:
\beq
\Lambda(-y)=-\Lambda(y) \qquad  A_\mu(-y) = -A_\mu(y) .
\label{gaugeparam}
\eeq
Thus the gauge transformation is forced to be trivial at the
$S^1/Z_2$ orbifold fixed points, $\Lambda(0)= 0$ and $\Lambda(\pi R)=0$,
and this is the crucial fact that will allow us to define gauge-invariant
Wilson {\it lines} that appear in the 4D effective theory as additional fields.

Considering scalar and fermion charged bulk fields $\Phi$ and $\Psi$
we are free to choose their $Z_2$ parities defined by $\Phi(-y)=\pm\Phi(y)$
for bosons and
\beq
\Psi(-y)=\pm \i \gamma_5 \Psi(y)
\label{fermipari}
\eeq
for fermions. Using the fact that $\i \gamma_5\Psi_{L,R}=\pm  \Psi_{L,R}$,
it is clear that even parity in Eq.(\ref{fermipari}) corresponds to
choosing $\Psi_L$ even and $\Psi_R$ odd and vice versa for odd
parity.  Consistency with Eq.(\ref{gaugeparam}) demands that the associated gauge
transformations and covariant derivatives for $\Phi$ and $\Psi$ are given by:
\beq
\Phi\rightarrow \mbox{exp}(-\i\epsilon(y) q\Lambda)\Phi \qquad
\Psi\rightarrow \mbox{exp}(-\i\epsilon(y) q\Lambda)\Psi
\label{gaugetrafo}
\eeq
and $D_M = \partial_M + iq \epsilon(y) A_M$, where $q$ is the $U(1)$ charge
of the respective bulk field, and the step function $\epsilon(y)$ is
\beq
\epsilon(y) = \left\{ \begin{array}{lll}   -1 & \rm{for} & -\pi R<y<0  \\
 0 & \rm{for} & y=0 \\
 +1 & \rm{for} & 0<y< \pi R \end{array} \right. \nonumber
\eeq
In summary, the parities for the bulk fields are:
\beq\label{parities}
\begin{array} {c|c|c|c|c|c}
     A_{\mu}  & A_5 & \Lambda & \Phi & \Psi_L & \Psi_R\\
    \hline
      -&+&-&\mbox{free}&+/-&-/+\\
\end{array}
\eeq
Because $\Lambda(0)=0=\Lambda(\pi R)$
any brane-localized fields on the orbifold branes at $y=0,\pi R$ 
are $U(1)$ gauge singlets.

The Wilson line of a gauge theory can be defined as the parallel
transport of the gauge connection along a path $C$ parameterized
by $s \in [0,1]$ from $x(0)=x_i$ to $x(1)=x_f$
and thus as a solution of
\begin{equation}\label{ptconstraint}
\frac{\d x^M}{\d s}D_M W(C)=0,
\end{equation}
where $D_M$ is the covariant derivative. According to this
definition, the Wilson loop transforms as
\begin{equation}
W(C)\rightarrow U(\Lambda(x_f))W(C)U^{-1}(\Lambda(x_i)).
\end{equation}
for a gauge transformation $\Lambda$. For a covariant derivative
defined as $D_M=\partial_M +iA_M$, the solution to
Eq.(\ref{ptconstraint}) is found to be
\begin{equation}\label{nswilson}
W(C)\equiv \exp\left( -\i\int_C \d s \dot{x}^M A_M \right) .
\end{equation}
A gauge invariant
operator can be defined from Eq.(\ref{nswilson}) on a general
manifold by choosing the path $C$ to be closed, yielding the well
known Wilson loop.  However on an orbifold
with a gauge parameter $\Lambda$ which vanishes at (certain) singularities
an open Wilson line linking those singularities is gauge invariant despite
the path $C$ not being closed.  In our $S^1/Z_2$ case we can 
define a gauge-invariant Wilson line phase $\Theta(x)$ by
\beq
e^{-\i\Theta(x)}\equiv \exp\left(-\i \int_{0}^{\pi R} A_5(x,y)\d y \right) .
\label{thetadef}
\eeq
As in the $M_4\times S^1$ case \cite{AHCCR},
the Wilson line phase $\Theta(x)$ implies the existence of
a scalar field of the 4D effective theory.  (Here we implicitly
make the assumption that the variation in $x$ is slow compared
to size $\pi R$ of the $S^1/Z_2$ orbifold direction, so that a
4D effective field theory description is valid.)  Note that $\Theta(x)$
is a {\it non-local} field from the 5D perspective.  By
calculating the propagators of charged bulk fields we show below how
the Wilson line phase can enter the 4D effective Lagrangian starting from
local 5D interactions.

The kinetic and interaction terms for $\Theta$ in the 4D effective action
follow from the 5D action, which has the form
\beq
S=\int \d^4x\int_{0}^{\pi R}\d y \biggl[ \mathcal{L}_{5D} +
\delta(yM_5)\mathcal{L}_0+\delta([y-\pi R]M_5)\mathcal{L}_{\pi R}\biggr]
\eeq
where the brane terms $\mathcal{L}_{0,\pi R}$ depend only on $Z_2$ even bulk fields and
brane localized fields, and $M_5$ is the cutoff scale of the 5D theory.  
The bulk action contains the gauge term
\beq
S_{5D,{\rm gauge}}=-\frac{1}{4g^2}\int\d^4x\d y F_{MN}F^{MN} +
\ldots \supset -\frac{2}{4g^2}\int\d^4x\d yF_{\mu 5}F^{\mu 5} .
\label{gaugeaction5d}
\eeq
To identify the 4D kinetic term for $\Theta$ following from this action
it is useful to choose a `uniform gauge' (constant in $y$)
representative of the gauge configuration $A_5$ with specified $\Theta(x)$ 
\beq\label{uniform}
A_5(x,y) \equiv A_5(x) = {\Theta(x)\over \pi R} .
\eeq
Then Eq.(\ref{gaugeaction5d}) implies a kinetic term for $\Theta$:
\beq
\int\d^4x\left(-\frac{1}{2\pi Rg^2}(\partial_{\mu}\Theta)^2\right)=
\int\d^4x\left(-\frac{1}{2}(\partial_\mu\sigma)^2\right).
\label{thetakin}
\eeq
where we have canonically normalized the 4D field by defining $\sigma(x)\equiv f\Theta(x)$ with
\beq
f^2={1\over \pi Rg^2} .
\eeq

Because of the gauge invariance of the original 5D action, and the non-local nature of $\Theta(x)$,
potential terms for the Wilson line phase, $\sigma(x)$, in the 4D action only arise from the propagation
of charged bulk fields from one brane to the other.  Thus we need to evaluate the propagator for  
charged bulk fields.  All our arguments can be illustrated using scalar fields and so the relevant matter
terms in the 5D bulk Lagrangian are
\beq
\mathcal{L}_{5D,{\rm matter}}=-(D_M\Phi)^{\dagger}(D^M\Phi)-m_{\Phi}^2\Phi^\dagger\Phi .
\eeq
The propagator is the Greens function for the equation of motion (with as usual a suitable Feynman $i\epsilon$ 
prescription imposed which we leave implicit).  We only need the propagator from
the $y=0$ brane to the $y=\pi R$ brane, but it is convenient to consider propagation
from the $y=0$ brane to an arbitrary position $y$ in the bulk.  Fourier transforming to momentum space
$p^\mu$ in the $x^\mu$ directions, but leaving the 5th direction in coordinate space, the
Greens function, $\Delta_{\Phi}(p,y)$, for a scalar in the slowly varying $A_5$
background solves
\beq
\left[ \partial_5^2-p^2+2\i q\epsilon(y)A_5\partial_5-q^2 A_5^2
-m_{\Phi}^2 \right] \Delta_{\Phi}(y)=-\delta(y) ,
\eeq
where we are here working in uniform gauge with $A_5(x,y)=A_5(x)$.
For the relevant case of even parity fields, Neumann boundary conditions
$\partial_5 \Delta_{\Phi}(y)=0$ have to be imposed at $y=0,\pi R$. 
The solution specialized to the case of interest, $y=\pi R$, is
\beq
\Delta_{\Phi,+}(p;0,\pi R) =\frac{p' \exp(-\i q A_5 \pi R)}{(p'^2+q^2A_5^2)\sinh(\pi R p')} .
\label{bosonproppiR}
\eeq
where $p'$ is given by $p'^2=p^2+m_{\Phi}^2$. 

The most important limit of the expression Eq.(\ref{bosonproppiR}) is when the separation between
the two branes is large compared to the inverse mass of the bulk particle, $\pi R m_{\Phi} \gg 1$:
\beq
\Delta_{\Phi,+}(p;0,\pi R) =\frac{2 e^{-\pi R m_{\Phi}}}{m_{\Phi}\left( 1+\frac{q^2 g^2 \sigma(x)^2}{\pi
R m_{\Phi}^2}\right)} \exp\bigl(-\i q \sqrt{g^2 \pi R} \sigma(x)\bigr) + {\mathcal O}(p^2),
\label{proppiR1}
\eeq
where we have switched to the canonically normalized 4D field $\sigma(x)$.  For values of
$\sigma$ where the phase is ${\mathcal O}(1)$ or less the condition $\pi R m_{\Phi} \gg 1$ implies
the further simplification
\beq
\Delta_{\Phi,+}(p;0,\pi R) =\frac{2 e^{-\pi R m_{\Phi}}}{m_{\Phi}}
\exp\bigl(-\i q \sqrt{g^2 \pi R} \sigma(x)\bigr) + {\mathcal O}(p^2) .
\label{proppiR2}
\eeq
Thus as claimed the propagation of charged bulk fields with even parity from
brane to brane leads to terms involving
the gauge-invariant Wilson-line phase
\beq
\exp\bigl(-\i q \sqrt{g^2 \pi R} \sigma(x)\bigr) .
\label{phase}
\eeq
As expected, in the limit that $\pi R \gg 1/m_{\Phi}$, the overall
amplitude of the Greens function is suppressed by $\exp(-m_{\Phi} \pi R)$.

Contributions to the 4D effective field theory involving the Wilson line phase
arise when the charged massive even-parity bulk fields which we integrate out
couple to brane-localized operators $j$ and $j'$ at $y=0$ and
$y=\pi R$:
\beq
S_{int}= \int\d^4x\int_{0}^{\pi R}\d y \biggl[ \lambda j(x)\Phi(x,y)\delta(yM_5)+
\lambda' j'(x)\Phi(x,y)\delta\bigl((y-\pi R)M_5\bigr)+\mbox{h.c.} \biggr]
\eeq
A tree level Feynman graph in the 5D theory coupling the charged bulk scalar to
currents on the branes then results to leading order and
for $m_{\Phi} \pi R\gg 1$ in an interaction in the 4D
effective theory given by
\beq
\lambda \lambda'^\dagger \frac{e^{-m_{\Phi}\pi R}}{m_{\Phi}}\int\d^4x j(x)
\e^{-\i q \sqrt{g^2 \pi R} \sigma(x)} j'^\dagger(x) + h.c.
\label{nonsusyint}
\eeq
The most important aspect of this non-derivative interaction of $\sigma$ is that it
is non-local from the 5D perspective, requiring the propagation of charged fields
across the 5D bulk.  This ensures that these interaction terms for $\sigma$ are
protected from large quantum gravity corrections, as these too must be non-local
in 5D, and are thus suppressed by the action of a black hole or wormhole or other
5D quantum gravitational effect of linear size $\pi R \gg 1/M_5$.   
Second, the size of the non-derivative interactions generated by charged bulk fields
is exponentially suppressed by the mass $m_{\Phi}$ for moderately large
bulk, so effects that lift the masslessness of $\sigma(x)$ can be
naturally exponentially suppressed.  Third, a true potential for $\sigma$
is generated if
both $j$ and $j'$ acquire vacuum expectation values.  This is not all that restrictive
as $j$ and $j'$ can be any operators on the $y=0,\pi R$ branes, as the $U(1)$ gauge
transformations vanish at $y=0,\pi R$ and arbitrary 4D Lorentz-scalar operators can
be coupled to $\Phi(x,0)$ and $\Phi(x,\pi R)$.

\section{Supersymmetric Wilson lines on \boldmath$S^1/Z_2$}

In the last Section we discussed the Wilson line on $S^1/Z_2$ and
its interactions with brane fields in the 4D effective theory
arising from 5D local interactions. In this Section we generalize
the discussion to a 5D $N=1$ supersymmetric theory compactified on
$S^1/Z_2$.  An $N=1$ 4D superfield description appropriate for such
theories is developed in Refs.\cite{5Dformalism,Hebecker} and we
here follow the conventions of \cite{Hebecker} and employ some
of the results of Ref.\cite{diplom}.

\subsection{The 4D \boldmath$N=1$ supersymmetric Wilson line}\label{ch5dmotivation}

For 4D $N=1$ supersymmetry, where a manifest superspace formulation is at hand,
the supersymmetric Wilson loop has been discussed previously in
\cite{4DWilson,Mkrtchian:1982us}.  The natural
generalization of the Wilson line in the supersymmetric case is
the solution to the parallel transport equation of the super gauge
connection along a path in superspace:
\beq
\frac{\d z^M}{\d s}\nabla_M W(C)=0 .
\label{sptconstraint}
\eeq
Here $z^M$ are superspace coordinates and $\nabla_M$ is the gauge
covariant derivative in superspace.
The formal solution to Eq.(\ref{sptconstraint}) is
\beq\label{susywilson}
W(C)\equiv \mathcal{P}\exp\left(-\int_{s_1}^{s_2} \d s \dot{z}^A
\mathcal{A}_A \right),
\eeq
where $\mathcal{A}_A$ is
the super gauge connection and $z^A=e^A_M z^M$ are the flat
superspace coordinates, where $e^A_M$ is the supervielbein.
In the conventions of Ref.\cite{WessBagger}, the supervielbein is
\beq
e^A_M=\left(
\begin{array}{ccc}
\delta^a_m & 0 & 0\\
-\i\sigma^a_{\mu\dot{\mu}}\thetab^{\mu} & \delta_{\mu}^{\alpha} & 0\\
-\i\theta^{\rho}\sigma^a_{\rho\dot{\nu}}\epsilon^{\dot{\nu}\dot{\mu}} & 0 & \delta^{\dot{\mu}}_{\dot{\alpha}}\\
\end{array}\right)
\label{superv}
\eeq
and the super gauge connection is given in terms of the fields $U$ and $V$ by
\beq\label{4DA}
\begin{split}
\mathcal{A}_{\alpha}&=-e^{2V}\D_{\alpha}e^{-2V}\\
\mathcal{A}_{\alphad}&=-e^{2U}\Dbar_{\alphad}e^{-2U}\\
\mathcal{A}_a&=\frac{\i}{4}\bar{\sigma}_a^{\betad\alpha}
\left[-\D_{\alpha}\mathcal{A}_{\betad}-\Dbar_{\betad}\mathcal{A}_{\alpha}+\{\mathcal{A}_{\alpha},\mathcal{A}_{\betad}\}\right]
\end{split}
\eeq
and the hermiticity constraint
$e^{2V}e^{-2U}=(e^{2V}e^{-2U})^{\dagger}$. Choosing $U\equiv 0$
just corresponds to a partial gauge fixing and defines the gauge
connection and hence the Wilson line purely in terms of the vector
superfield $V$. Under a super gauge transformation, the solution
Eq.(\ref{susywilson}) transforms as
\beq
W(C)\rightarrow U(\Lambda(z_f))W(C)U^{-1}(\Lambda(z_i)),
\eeq
where now $\Lambda$ is a chiral super field.

\subsection{The \boldmath$S^1/Z_2$ supersymmetric Wilson line}\label{ch5DV}

We are, however, interested in the Wilson line in 5D $N=1$ supersymmetric
theories on $S^1/Z_2$ in the case that the 4D $U(1)$ vector multiplet
is projected out by the orbifold conditions.  This changes the above construction
in a manner we now explain.  We follow the conventions of \cite{Hebecker}.

Upon dimensional reduction on a circle, 5D
$N=1$ supersymmetry leads to 4D $N=2$ supersymmetry with an
$SU(2)_R$ symmetry relating the two superfield transformations and
the central charge $p_5$.  Only one of these supersymmetries is
manifest in the 4D superfield formulation of 5D, $N=1$ supersymmetry
in terms of 4D, $N=1$ superfields in Wess-Zumino gauge of Refs.\cite{5Dformalism, Hebecker}.
Fortunately, on $S^1/Z_2$, the conservation of the fifth component 
of momentum, $p_5$, is
broken by the orbifold projection, breaking 4D $N=2$ down to $N=1$, and the
surviving $N=1$ can be chosen as the manifest supersymmetry in
the 4D superfield formulation.  Given these facts we can employ the result
Eq.(\ref{susywilson}) if we are able to identify the correct replacement
for $V$ in terms of the original 5D field content.

The field content of a 5D $N=1$ vector supermultiplet is a
vector-field $v^M$, a real scalar $\Sigma$ and an $SU(2)_R$ gaugino
doublet $\lambda^i_\alpha$, to which an $SU(2)_R$ triplet of real
auxiliary fields is added for the off-shell multiplet.
The off-shell action is given by 
\beq\label{componentaction}
S=\int \d^4x\:\d y\frac{1}{g^2}\Big\{-\frac{1}{4}(F_{MN})^2-
\frac{1}{2}(\partial_M\Sigma)^2-\frac{1}{2}\bar{\lambda}_i\i\Gamma^M\partial_M\lambda^i
+\frac{1}{2}(X^a)^2\Big\}
\eeq
where $F_{MN}$ is the field strength tensor.

As has been shown in \cite{Hebecker}, the supersymmetry
transformations remaining after orbifolding are the supersymmetry
transformations of the components of the vector superfield\footnote{In order to get a
non-vanishing Wilson line, we choose opposite $Z_2$ parities
for gauge fields and supergauge transformations compared to \cite{Hebecker},
and thus we modify definitions accordingly as we go along.}
\beq\label{defV5}
  V=-\theta\sigma^m\thetab v_m +\i\theta^2\thetab\bar{\lambda}_L-\i\thetab^2\theta\lambda_L
  +\frac{1}{2}\theta^2\thetab^2(X^3-\partial_5\Sigma)
\eeq
with the super gauge transformation
\beq
  V\rightarrow V +\frac{1}{2}(\Lambda+\Lambdad)
\eeq
and the chiral superfield
\beq\label{defPhi5}
  \Phi=(\Sigma+\i A_5)+\sqrt{2}\theta(-\i\sqrt{2}\lambda_R)+ \theta^2(X^1+\i X^2),
\eeq
if $\Phi$ has the inhomogeneous super gauge transformation
\beq\label{Phitrafo}
  \Phi\rightarrow \Phi+\partial_5\Lambda.
\eeq
This $\Phi$ supergauge transformation is necessary to stay
in Wess-Zumino gauge, see Ref.\cite{Hebecker}.

Now, Eq.(\ref{Phitrafo}) implies that for
\emph{odd} $\Phi$, $\nabla_5\equiv(\partial_5+\Phi)$ ought to be
interpreted as a covariant derivative, acting on
a chiral superfield in the $y$-direction.  For our case of \emph{even}
$\Phi$, conservation of the $Z_2$ parity of the chiral superfield, of $U(1)$
charge $q$ upon which the covariant derivative acts requires the modified definition 
\beq
\nabla_5\equiv(\partial_5+q\epsilon(y)\Phi).
\label{nablachiral}
\eeq
The covariant derivative acting on the real superfield $V$ is however
\beq
\nabla_5 V\equiv
\partial_5 V-\frac{1}{2}(\Phi+\Phi^{\dagger})
\eeq
independent of the parity choice.  The superfield action
\beq\label{5DSYMac}
S=\int\d^8z\:\d y \frac{1}{g^2}\left[\frac{1}{4}\left(W^\alpha
W_\alpha\delta^2(\thetab)+\text{h.c.}\right)+(\nabla_5 V)^2\right]
\eeq
reproduces the component field action Eq.(\ref{componentaction}) up to
irrelevant surface terms with the supersymmetric field
strength, $W_\alpha\equiv -\frac{1}{4}\Dbar^2D_\alpha V$.

We are now ready to identify the vector superfield gauge
connection and, via Eq.(\ref{susywilson}), the supersymmetric Wilson
line phase and its kinetic term.  Since the full 5D (super)Poincare group is broken
by the orbifold boundary conditions it is convenient to think of $y$ as just a
parameter of the theory, allowing us to work in terms of 4D superfields. 
We only need to consider the $N=1$ supersymmetry remaining after
the $S^1/Z_2$ compactification, so the extension of the 4D
supersymmetric vielbein of Eq.(\ref{superv}) to our case simply involves, 
allowing $x$-space curved and tangent space indices, $a$ and $m$, to take on
the values $0,1,2,3,5$.  In fact for the paths of interest which
stretch from one orbifold brane to the other with
constant $\theta$ and $\thetab$, we can forget about the vielbein
in the definition Eq.(\ref{susywilson}).

The parallel transport of the Wilson line in the $y$ direction requires
\beq
\frac{\d y}{\d s}\nabla_5 W(C_5)=0,
\eeq
and from the gauge transformation of $\Phi$ in Eq.(\ref{Phitrafo}), it is
clear that this is satisfied by
\beq
W(C_5)\equiv \exp\left(-\int_{s_1}^{s_2} \d s \frac{\d y}{\d s}
\Phi\right)
\eeq
where the sign in the exponent is fixed by demanding the
gauge transformation
\beq
W(C_5)\rightarrow \e^{-\epsilon(y)\Lambda(y_f)}W(C)e^{\epsilon(y)\Lambda(y_i)} .
\eeq
In other words we identify the supergauge connection in
$y$-direction to be
\beq
\mathcal{A}_5\equiv \Phi.
\eeq
and the supersymmetric Wilson line 
\beq
W(C_5)\equiv e^{-\Theta} \equiv \exp\left(-\int_0^{\pi R}\d y \Phi \right) .
\label{susywilson5}
\eeq
The form of supersymmetric Wilson line in terms of component
fields can be obtained from Eq.(\ref{susywilson5}) by expanding out
$W(C)$ in a power series of $\theta$ and $\thetab$.  From the component expansion of
$\Phi$ in Eq.(\ref{defPhi5}) one can see that the $\theta$ and
$\thetab$-independent pure-phase part of this expression
reproduces the expected Wilson line phase associated to $A_5$.
Analogous to the non-supersymmetric Wilson line, we interpret the
supersymmetric Wilson line to be the exponential of a chiral (pseudo)Goldstone
multiplet $\Theta$. 

The kinetic term of this multiplet arises
from the original 5D bulk vector multiplet action.  Explicitly, 
taking $\Phi$ to be in the `uniform' gauge, i.e. setting
\beq
\Phi(x,y)=\Phi(x,0),
\eeq
independent of $y$, substituting this into Eq.(\ref{5DSYMac}) and integrating
out $y$, we obtain the kinetic term for $\Theta$:
\beq
\int \d^8z f^2\Theta^{\dagger}\Theta + h.c.\equiv\int \d^8z \sigma^{\dagger}\sigma + h.c.
\eeq
with $\sigma=f\Theta$ the canonically normalized 4D field
where $f^2=2/\pi R g^2$ similar to the non-supersymmetric case.

The most important feature of the supersymmetric Wilson-line, Eq.(\ref{susywilson5})
is that it not only contains the real pseudoscalar field $A_5$ (or more precisely the 
pseudoscalar component of $W(C_5)$) but also the superpartners, the scalar
$\Sigma$ and the Weyl fermion $\lambda_R$ as seen in the component expansion
of $\Phi$, Eq.(\ref{defPhi5}).  One possible application of this multiplet
is as an `axion' multiplet, so we will sometimes refer to the pseudoscalar, scalar,
and Weyl fermion components of $W(C_5)$ as the axion, saxion, and
axino, respectively, though in this paper
we will not investigate such supersymmetric axion phenomenology in
detail (see Refs.\cite{Choi} and \cite{DDG}
for work along these lines in the non-supersymmetric context).  As we discuss in Section
\ref{kkmodes} although the 4D field $W(C_5)$ originates from a 5D field it has the
interesting property that it does not possess physical Kaluza-Klein mode excitations
as one might naively expect.

\subsection{4D Wilson line interactions from local 5D interactions}

As we will show in this section, non-derivative interactions for the $\Theta$
supermultiplet arise in the effective 4D theory from charged 5D bulk
hypermultiplets propagating between the branes in analogy to the
non-supersymmetric Wilson line.  For the general discussion of the
hypermultiplet we again follow
\cite{5Dformalism,Hebecker}. We again specialize to $U(1)$ and modify
the expressions according to our $Z_2$ parities.

The field content of a hypermultiplet is given by a complex scalar
$SU(2)_R$ doublet $H^i$ of opposite $Z_2$ parity and an $SU(2)_R$ singlet Dirac spinor
$\Psi$ which can be decomposed into two Weyl spinors
$\Psi=(\psi,\bar{\psi}^c)^T$ of opposite $Z_2$ parity.  A complex scalar $SU(2)_R$
doublet $\tilde{F}_i$ of auxiliary fields of opposite $Z_2$ parity is added for the
off-shell multiplet.  The supersymmetric action is 
\beq
  S=\int\d^4x\:\d y\left( -(\partial_M H_i)^{\dagger}(\partial^MH^i)-
  \bar{\Psi}(\i\Gamma^M\partial_M+\epsilon(y)m)\Psi\right).
\eeq
Choosing the $Z_2$ parities as 
\beq\label{Hparities}
\begin{array} {c|c|c|c|c|c}
     H^1  & \psi_L & F_1 & H^2 & \psi_R & F_2\\
    \hline
      +&+&+&-&-&-\\
\end{array},
\eeq
the supersymmetry transformations of the component fields can be
identified as the superfield transformations of the chiral
superfields 
\beq\label{defH}
  \begin{split}
    H=&H^1+\sqrt{2}\theta\psi_L+\theta^2(F_1+\partial_5H^2)\\
    H_c=&H^{\dagger}_2+\sqrt{2}\theta\psi_R+\theta^2(-F^{2 \dagger}-\partial_5H^{\dagger}_1),
  \end{split}
\eeq
and the 5D action in terms of these superfields is 
\beq
  S=\int \d^8z\:\d y \left\{\Hd H+H_c \Hd_c+\left(H_c(\partial_5+\epsilon(y)m)H\delta(\thetab)+\mbox{h.c.}\right)\right\} .
\eeq

The next step is to couple the hypermultiplet to the vector
supermultiplet.  Let $H$ carry $U(1)$ charge $q$:
\beq
H \rightarrow e^{-q\epsilon(y)\Lambda}H , \qquad H_c \rightarrow H_c e^{{q\epsilon(y)\Lambdad}}
\eeq
where, as $\Lambda$ is odd, we have to introduce $\epsilon(y)$.
Then the action for the hypermultiplet becomes
\beq\label{5Dchint}
S=\int\d^8z\:\d y \left\{\Hd e^{2q\epsilon(y)V}H+H_c e^{-2q\epsilon(y)V}\Hd_c
+\left(H_c(\nabla_5+\epsilon(y)m)H\delta(\thetab)+\text{h.c.}\right)\right\} 
\eeq
where
\beq
\nabla_5H\equiv(\partial_5+q\epsilon(y)\Phi)H.
\eeq 
defines the action of the covariant derivative.
There is one subtlety in the above treatment:\footnote{See \cite{Hebecker} and \cite{diplom} for
additional discussion in the case of $Z_2$ gauge supermultiplet parities opposite to ours.}
in gauging the Hypermultiplet, the definitions of the auxiliary fields have to be modified to
\beq\label{defH2}
  \begin{split}
    H=&H^1+\sqrt{2}\theta\psi_L+\theta^2(F_1+D_5H^2-\epsilon(y)\Sigma)\\
    H_c=&H^{\dagger}_2+\sqrt{2}\theta\psi_R+\theta^2(-F^{2 \dagger}-D_5H^{\dagger}_1-\epsilon(y)H^{\dagger}_1\Sigma).
  \end{split}
\eeq 

From Eq.(\ref{5Dchint}), we derive the superfield propagators in
Appendix \ref{appPhiprop}.  Note that, as indicated in the last
chapter, we treat $\Phi(x,0)$ as a background field and thus
expect a dependence of the hypermultiplet propagator on
$\Phi(x,0)$.  The result for the hypermultiplet propagator
is:\footnote{As in the non-supersymmetric case, we fix one end
of the propagator on the $y=0$ brane.}
\begin{equation}
\Delta_H=\left(
	\begin{array}{cc}
		(-\partial_5+\epsilon(y)(m+\Phi))\frac{\Dbar^2}{4\Box} & \frac{\Dbar^2\D^2}{16} \\
		\frac{\D^2\Dbar^2}{16} & (\partial_5+\epsilon(y)(m+\Phid))\frac{\D^2}{4\Box}\\
	\end{array}
	\right)
	\left(
	\begin{array}{cc}
		\mathcal{G}_1(y) & 0 \\
		0 & \mathcal{G}_2(y)\\
	\end{array}
	\right)
	\delta^4(\theta-\theta')
\end{equation}
where the $2\times 2$ matrix indicates the $(H_c,H^{\dagger})\times (H, H^{\dagger}_c)$
propagators. Imposing boundary conditions according to  $H$ being $Z_2$ even and hence
$H_c$ being odd, we calculate the full $y$ dependent $\mathcal{G}_{1,2}$ in the appendix
(see Eq.(\ref{Gfullsoln})). Note however, that the potential for the Wilson line from
the coupling of the hypermultiplet to brane localized fields at $y=0$ and $y=\pi R$ only
depends on $\frac{1}{16}\Dbar^2 \D^2 \mathcal{G}_2(\pi R)$ corresponding to the
$H^{\dagger}H$ propagator as $H_c$ (and $H_c^{\dagger}$) are $Z_2$ odd.
For the brane-to-brane case of interest, $y=\pi R$, and $\mathcal{G}_2$ reads
\beq\label{Gdef}
  \mathcal{G}_2  =  \frac{1}{\mathcal{N}'_2}e^{-\frac{q}{2}(\Phi-\Phid)\pi R}
\eeq
where
\beq
\mathcal{N}'_2 \equiv \frac{2\sinh(p'\pi R)}{p'}(p^2+|m+q\Phi|^2)
\eeq
and
\beq
p'^2 \equiv p^2+\left(m+\frac{q}{2}(\Phi+\Phid)\right)^2.
\eeq

In a similar fashion to the non-supersymmetric case this brane-to-brane propagator
may be simplified in various limits, the most important of which is the case
where the inter-brane separation is large in units of the bulk hypermultiplet mass.
As for the non-supersymmetric Wilson line phase this results in the interactions
of the `axion' and its superpartners with brane-localized fields being exponentially
suppressed by $\exp(-m\pi R)$.  This is of importance for applications
of this mechanism to model-building where it is desired that the masses and non-derivative
interactions of the pseudo-Goldstone multiplet are suppressed relative to the 
fundamental scale $M_5$, as well as being protected from possible global-symmetry-violating
quantum gravitational effects.

\subsection{Kaluza-Klein modes and the Wilson line}\label{kkmodes}

By using the $y$-dependent superfield formalism no expansion and resummation in Kaluza-Klein
(KK) modes was needed to achieve our results.  However, to make contact to the KK
picture and clarify a few subtleties related to it, a few comments are in order.
For the non-supersymmetric Wilson line discussed in Section \ref{flat} we calculated the Wilson
line kinetic term and its potential in uniform gauge Eq.(\ref{uniform}). As indicated, this is
justified as any local gauge-field contribution can be gauged away.  However the brane-to-brane
Wilson line is gauge invariant and therefore not all information contained in the $A_5$ gauge field
can be entirely gauged away on $S_1/Z_2$.  In particular the
$y$-independent mode of $A_5$ is physical.\footnote{Another
way to see this is that the gauge transformation needed to gauge away the constant field
is $\Lambda=y$ which does not satisfy the boundary conditions on $S_1/Z_2$.} Therefore, in terms of
KK modes, the pseudo scalar field given by the Wilson line phase contains only a
single physical zero mode while {\it all higher KK-modes
correspond to unphysical gauge degrees of freedom.}

In the supersymmetric case the Wilson line multiplet
contains the Wilson line phase as its imaginary scalar
component.  As only the zero mode is a physical mode of the scalar component (by the previous 
gauge transformation argument), this must be true for the whole
supermultiplet too.  However, as for the treatment in 4D superfields it was necessary to work in
Wess-Zumino gauge Eq.(\ref{Phitrafo}). Therefore, all components of the
super gauge transformation $\Lambda$ apart from the one we use to fix uniform gauge for the
$A_5$ component are fixed by Wess-Zumino gauge condition and cannot be used to gauge away higher
KK-modes of $\Sigma$ and $\lambda_R$.  Instead in Wess-Zumino gauge the higher KK modes of
$\Sigma$ and $\lambda_R$ are ``eaten'' by the KK-modes of the massive vector supermultiplet
$V$, providing the bosonic longitudinal degree of freedom for the  $A_{\mu}^{(n)}$ KK-modes
and the necessary fermionic degrees of freedom for the gauginos $\lambda_L$ to form
massive KK Dirac fermions.  Thus the Wilson line supermultiplet
arising from a 5D vector supermultiplet on $S^1/Z_2$ provides a 4D chiral pseudo-Goldstone
supermultiplet with a zero mode and {\it no higher KK-modes}.

\section{Conclusions}
In this paper we considered a $U(1)$ gauge theory on the five dimensional orbifold
$\mathcal{M}_4\times S^1/Z_2$, where $A_5$ has even $Z_2$ parity.  We
showed that this leads to a light pseudoscalar degree of freedom $W(x)$
in the effective 4D theory below the compactification scale
arising from a gauge-invariant brane-to-brane Wilson line.
As noted by Arkani-Hamed \etal in the non-supersymmetric $S^1$ case the 5D bulk
gauge-invariance of the underlying theory together with the non-local
nature of the Wilson line field leads to the protection of the
4D theory of the Wilson line field $W(x)$ from possible large global-symmetry
violating quantum gravitational effects.  We studied the $S^1/Z_2$
theory in detail, in particular developing the supersymmetric generalization
of this construction, involving a pseudoscalar Goldstone field (the `axion')
and its scalar and fermion superpartners (`saxion' and `axino').  The
global nature of $W(x)$ implies the absence of independent
Kaluza-Klein excitations of its component fields.  The non-derivative
interactions of the Wilson line degree of freedom in
the effective 4D theory arise from $U(1)$ charged 5D fields $\Phi$ propagating
between the boundary branes.  Because such effects are suppressed by
$\exp(-\pi R m_\Phi)$ a small hierarchy between the
inverse mass $1/m_\Phi$ of such bulk scalars and the size, $\pi R$, of the
5th dimension leads to an exponentially large suppression of the
non-derivative couplings of the Wilson line.  Thus the
pseudo-Goldstone nature of the 4D Wilson line field is easy to
maintain against quantum gravity effects, and quite naturally the
pseudo-Goldstone field (and it's superpartners) can have mass much
smaller than the fundamental scales, $M_5$, or $1/R$ of the theory.
Given these noteworthy properties we believe that
investigation of the model building uses of such Wilson
line degrees of freedom arising from $A_5$ zero modes
on $\mathcal{M}_4\times S^1/Z_2$ and
$\mathcal{M}_4\times S^1/(Z_2\times Z_2')$ orbifolds
is warranted.

\section*{Acknowledgements}
T.F. wishes to thank the Evangelisches Studienwerk e.V. Villigst and PPARC
for stipendiary grants. B.H.'s work was supported by the Clarendon Fund.

\appendix
\section{Conventions and notation}\label{appconv}
We use the metric $\d s^2=\eta_{MN}\d x^M\d x^N=\eta_{\mu\nu}\d
x^{\mu}\d x^{\nu}+\d y^2$, where $\eta_{\mu\nu}=\mbox{diag}(-1,1,1,1)$, $M,N=0,1,2,3,5$ and $\mu,
\nu=0,1,2,3$ and $0\leq y \leq \pi R$ is the fundamental domain
after $S^1/Z_2$ compactification. When dealing with fermions, we
use the following representation of the $\gamma$ matrices:
\begin{equation}
  \gamma_\mu=\left(
    \begin{array}{c c}
      0 & \sigma_{\mu}\\
      \bar{\sigma}_{\mu} & 0 \\
    \end{array}\right),
  \gamma_5=\left(
    \begin{array}{c c}
      -i & 0\\
      0 & i \\
    \end{array}\right),
\end{equation}
where $\sigma_\mu=(1_2,\sigma_i)$,
$\bar{\sigma}_\mu=(1_2,-\sigma_i)$ and $\sigma_i$ are the
$\sigma$ matrices.

To fix our conventions for the 5D super gauge transformations, we
choose
\begin{equation}\label{gaugetrafodef}
\e^{\epsilon(y)2V}\rightarrow \e^{\epsilon(y)\Lambdad}\e^{\epsilon(y)2V}\e^{\epsilon(y)\Lambda}
\end{equation}
which agree with the conventions of \cite{Hebecker} apart from the step functions which we introduce as we choose $V$, $\Lambda$ to have odd $Z_2$ parity.

Via Eq.(\ref{4DA}), Eq.(\ref{gaugetrafodef}) yields
\begin{equation}
\mathcal{A}_M\rightarrow
e^{-\epsilon(y)\Lambda}(\mathcal{A}_M+\partial_M)e^{\epsilon(y)\Lambda}.
\end{equation}
From this, the super gauge transformation for a Wilson line of a
path with constant $\theta,\thetab$ is
\begin{equation}
W(C)\rightarrow \e^{-\epsilon(y)\Lambda(x_f)}W(C)e^{\epsilon(y)\Lambda(x_i)}.
\end{equation}

\section{Calculation of superfield hypermultiplet propagator}\label{appPhiprop}

Expanding the Hypermultiplet super-action Eq.(\ref{5Dchint}) leads to
\begin{equation}\label{5Dchexp}
  \begin{split}
    S=\int\d^8z\:\d y &\left\{ \Hd H+H_c \Hd_c +H_c(\partial_5+\epsilon(y)m+q\epsilon(y)\Phi)H\delta(\thetab)\right.\\
	&\left.\:+\Hd(-\partial_5+\epsilon(y)m+q\epsilon(y)\Phi^{\dagger})\Hd_c\delta(\theta)+O(V)\right\}\\
  \end{split}
\end{equation}

As for the non-supersymmetric theory, we consider the propagation of the hypermultiplet in the presence of
a $\Phi(x,0)$ background field. Thus we obtain the following free
generating functional\footnote{We denote $\J(z,y)\equiv \J$ and
$\J(z',y')\equiv \J'$.}
\begin{equation}
  \begin{split}
    Z_{0,H}(\J_H,\Jd_H,\J_{H_c},\Jd_{H_c})=&\int \mathcal{D}H\mathcal{D}\Hd\mathcal{D}H_c\mathcal{D}\Hd_c\\
    &\exp\left\{\i\int\d^8z\:\d y \left[(H_c, \Hd)M_H{H \choose \Hd_c}\right.\right.\\
      &\left.\left.\qquad+(H_c, \Hd)
        \left(\begin{array}{cc}
            -\frac{\D^2}{4\Box} & 0\\
            0 & -\frac{\Dbar^2}{4\Box}
          \end{array}\right)
        {\J_{H_c} \choose \Jd_H}
      \right.\right.\\
      &\left.\left.\qquad
        +(H, \Hd_c)
        \left(\begin{array}{cc}
            -\frac{\D^2}{4\Box} & 0\\
            0 & -\frac{\Dbar^2}{4\Box}
          \end{array}\right)
        {\J_{H} \choose \Jd_{H_c}}\right]\right\}\\
=\exp\left\{-\i\int \d^8\right.&\left.z\:\d
y(\J_{H_c},\Jd_{H})_i{\Delta_{5,H}}_{ij}(z-z',y-y'){\J'_{H}\choose
\J'^{\dagger}_{H_c}}\right\}
\end{split}
\end{equation}
where $M_H$ is
\begin{equation}\label{defM}
(M_H)= \left(\begin{array}{cc}
                      (\partial_5+\epsilon(y)m+\epsilon(y)q\Phi)(-\frac{\D^2}{4\Box}) & 1\\
                      1 & (-\partial_5+\epsilon(y)m+\epsilon(y)q\Phid) (-\frac{\Dbar^2}{4\Box})
                      \end{array}\right)
\end{equation}
and the chiral propagator $(\Delta_{5,H})$ is defined by
\begin{equation}\label{defprop}
  \left(\begin{array}{cc}
      \frac{\Dbar^2}{4\Box} & 0\\
      0 & \frac{\D^2}{4\Box}
    \end{array}\right)
  (M_H)(\Delta_{5,H})\equiv
  - \left(\begin{array}{cc}
      - \frac{\Dbar^2}{4\Box} & 0\\
      0 & - \frac{\D^2}{4\Box}
    \end{array}\right)
  \delta^8(z-z')\delta(y-y').
\end{equation}

To solve for the propagator we invert $(M_H)$, using the method of
supersymmetry projectors outlined in \cite{WessBagger}. From Eq.(\ref{defM}):
\beq
   M_H=\left(\begin{array}{cc}
	-\Box^{-\frac{1}{2}}bP_+ & 1\\
	1    & -\Box^{-\frac{1}{2}} cP_-\\
	\end{array}\right)
\eeq
where
\bea
      b & = & \partial_5+\epsilon(y)m+\epsilon(y)q\Phi\\
      c & = &-\partial_5+\epsilon(y)m+\epsilon(y)q\Phid.
\eea
Inverting this via supersymmetry projector algebra and using it to solve Eq.(\ref{defprop}) yields
\beq
  \Delta_H=\left(\begin{array}{cc}
	\frac{c}{\Box-bc}\frac{\Dbar^2}{4} & \frac{1}{\Box-cb} \frac{\Dbar^2\D^2}{16}\\
	\frac{1}{\Box-bc}\frac{\D^2\Dbar^2}{16} & \frac{b}{\Box-cb}\frac{\D^2}{4}\\
        \end{array}\right)
  \delta(z-z')\delta(y-y')
\eeq
up to terms which vanish under a super-integral under which $\Delta_H$ is defined.
 
Fourier transformation in the non compactified directions and rewriting this leads to
\beq
  \Delta_H=\left(\begin{array}{cc}
	c\frac{\Dbar^2}{4} &\frac{\Dbar^2\D^2}{16}\\
	\frac{\D^2\Dbar^2}{16} & b\frac{\D^2}{4}\\
        \end{array}\right)
  \left(\begin{array}{cc}
  	\mathcal{G}_1& 0 \\
	0 & \mathcal{G}_2\\
        \end{array}\right)
  \delta^4(\theta-\theta')
\eeq
where $\mathcal{G}_{1,2}$ are solutions to
\bea
    (-p^2-bc)\mathcal{G}_1 & = & \delta(y-y')\\
    (-p^2-cb)\mathcal{G}_2 & = & \delta(y-y')
\eea
and subject to the appropriate boundary conditions.

As in the non-susy case, to calculate the propagator, we fix
$y'=0$. The equations read
\bea\label{susyde}
    \Big[\partial_5^2+\epsilon(y)q(\Phi-\Phid)\partial_5-2\delta(y)(m+q\Phid) - p^2 - (m+q\Phi)(m+q\Phid)\Big]\mathcal{G}_1 & = & \delta(y)\\
    \Big[\partial_5^2+\epsilon(y)q(\Phi-\Phid)\partial_5+2\delta(y)(m+q\Phi) - p^2 - (m+q\Phi)(m+q\Phid)\Big]\mathcal{G}_2 & = &\delta(y).
\eea

As $H$ is even and $H_c$ is odd, we have to chose even and odd boundary
conditions for $\mathcal{G}_2$ and $\mathcal{G}_1$ respectively. With
these boundary conditions, the solutions to Eq.(\ref{susyde}) are

\bea\label{Gfullsoln}
     \mathcal{G}_1 & = & e^{-\frac{q}{2}(\Phi-\Phid)|y|}\frac{1}{\mathcal{N}_1}\sinh(p'(|y|-\pi R))\\
     \mathcal{G}_2 & = & e^{-\frac{q}{2}(\Phi-\Phid)|y|}\frac{1}{\mathcal{N}_2}\left(p'\cosh(p'(|y|-\pi R))+\frac{q}{2}(\Phi-\Phid)\sinh(p'(|y|-\pi R))\right)
\eea
where
\begin{equation}
p'^2\equiv p^2+\left(m+\frac{q}{2}(\Phi+\Phid)\right)^2
\end{equation}
and the normalization factors are
\bea
     \mathcal{N}_1 & \equiv & q(\Phi-\Phid)\sinh(p'\pi R)-2p'\cosh(p'\pi R)\\
     \mathcal{N}_2 & \equiv & 2\sinh(p'\pi R)\left(p^2+|m+q\Phi|^2\right).
\eea
Evaluating $\mathcal{G}_2$ at $y=\pi R$ and absorbing $p'$ into the normalization then leads to the result given in Eq.(\ref{Gdef}).


\begin{thebibliography}{99}

\bibitem{orbguts}
Y.~Kawamura,
Prog.\ Theor.\ Phys.\  {\bf 105} (2001) 999
[arXiv:hep-ph/0012125]:\\
G.~Altarelli and F.~Feruglio, Phys.\ Lett.\ B {\bf 511} (2001) 257
[arXiv:hep-ph/0102301].

\bibitem{orbgutsunif}
L.~J.~Hall and Y.~Nomura, Phys.\ Rev.\ D {\bf 64} (2001) 055003
[arXiv:hep-ph/0103125]:\\
A.~Hebecker and J.~March-Russell, Nucl.\ Phys.\ B {\bf 613} (2001) 3
[arXiv:hep-ph/0106166]; and
Nucl.\ Phys.\ B {\bf 625} (2002) 128 [arXiv:hep-ph/0107039].


\bibitem{su5}
A.~B.~Kobakhidze, Phys.\ Lett.\ B {\bf 514} (2001) 131 
[arXiv:hep-ph/0102323];\\
R.~Barbieri, L.~J.~Hall and Y.~Nomura, Nucl.\ Phys.\ B {\bf 624} (2002) 63\\{}
[arXiv:hep-th/0107004];\\
J.~A.~Bagger, F.~Feruglio and F.~Zwirner, Phys.\ Rev.\ Lett.\  {\bf 88} 
(2002) 101601 [arXiv:hep-th/0107128];\\
T.~j.~Li, Phys.\ Lett.\ B {\bf 520} (2001) 377 [arXiv:hep-th/0107136].

\bibitem{so10}
T.~Asaka, W.~Buchm\"uller and L.~Covi, Phys.\ Lett.\ B {\bf 523} (2001) 
199\\{} [arXiv:hep-ph/0108021]
and arXiv:hep-ph/0204358;\\
L.~J.~Hall \etal, Phys.\ Rev.\ D {\bf 65} (2002) 035008
[arXiv:hep-ph/0108071];\\
R.~Dermisek and A.~Mafi, Phys.\ Rev.\ D {\bf 65} (2002) 055002 
[arXiv:hep-ph/0108139].

\bibitem{CPRT}
R.~Contino, \etal, Nucl.\ Phys.\ B {\bf 622} (2002) 227
[arXiv:hep-ph/0108102];\\
Y.~Nomura, Phys.\ Rev.\ D {\bf 65} (2002) 085036 [arXiv:hep-ph/0108170].

\bibitem{HMROS}
L.~Hall, \etal
JHEP {\bf 0409}, 026 (2004) [arXiv:hep-ph/0108161];\\
L.~J.~Hall, Y.~Nomura and D.~R.~Smith, arXiv:hep-ph/0107331;\\
N.~Haba \etal, Prog.\ Theor.\ Phys.\  {\bf 107} (2002) 151 
[arXiv:hep-ph/0107190];\\
H.~D.~Kim, J.~E.~Kim and H.~M.~Lee, arXiv:hep-ph/0112094;\\
A.~Hebecker and J.~March-Russell,
Phys.\ Lett.\ B {\bf 541}, 338 (2002) [arXiv:hep-ph/0205143].

\bibitem{AHCCR}
N.~Arkani-Hamed, H.~C.~Cheng, P.~Creminelli and L.~Randall,
Phys.\ Rev.\ Lett.\  {\bf 90}, 221302 (2003)
[arXiv:hep-th/0301218]:
JCAP {\bf 0307}, 003 (2003)
[arXiv:hep-th/0302034].

\bibitem{quantgrav}
L.~M.~Krauss and F.~Wilczek,
Phys.\ Rev.\ Lett.\  {\bf 62}, 1221 (1989):\\
M.~G.~Alford, J.~March-Russell and F.~Wilczek,
Nucl.\ Phys.\ B {\bf 337}, 695 (1990):\\
M.~Kamionkowski and J.~March-Russell,
Phys.\ Rev.\ Lett.\  {\bf 69}, 1485 (1992)
[arXiv:hep-th/9201063];
Phys.\ Lett.\ B {\bf 282}, 137 (1992)
[arXiv:hep-th/9202003]:\\
R.~Holman, \etal,
Phys.\ Rev.\ Lett.\  {\bf 69}, 1489 (1992);
Phys.\ Lett.\ B {\bf 282}, 132 (1992)
[arXiv:hep-ph/9203206].

\bibitem{Kaplan}
D.~E.~Kaplan and N.~J.~Weiner,
JCAP {\bf 0402}, 005 (2004)
[arXiv:hep-ph/0302014].

\bibitem{Hofmann}
R.~Hofmann, F.~Paccetti Correia, M.~G.~Schmidt and Z.~Tavartkiladze,
Nucl.\ Phys.\ B {\bf 668}, 151 (2003)
[arXiv:hep-ph/0305230].

\bibitem{Pilo}
L.~Pilo, D.~A.~J.~Rayner and A.~Riotto,
Phys.\ Rev.\ D {\bf 68}, 043503 (2003)
[arXiv:hep-ph/0302087].


\bibitem{Choi}
K.~W.~Choi,
Phys.\ Rev.\ Lett.\  {\bf 92} (2004) 101602
[arXiv:hep-ph/0308024].



\bibitem{5Dformalism}
N.~Marcus, A.~Sagnotti and W.~Siegel,
Nucl.\ Phys.\ B {\bf 224} (1983) 159,
E.~A.~Mirabelli and M.~E.~Peskin,
Phys.\ Rev.\ D {\bf 58} (1998) 065002
[arXiv:hep-th/9712214],
N.~Arkani-Hamed, T.~Gregoire and J.~Wacker,
JHEP {\bf 0203} (2002) 055
[arXiv:hep-th/0101233].


\bibitem{Hebecker}
A.~Hebecker,
Nucl.\ Phys.\ B {\bf 632} (2002) 101
[arXiv:hep-ph/0112230].

\bibitem{diplom}
T.~Flacke,
DESY-THESIS-2003-047


\bibitem{4DWilson}
S.~J.~J.~Gates,
Phys.\ Rev.\ D {\bf 16} (1977) 1727.
S.~Marculescu and L.~Mezincescu,
Nucl.\ Phys.\ B {\bf 181} (1981) 127.
R.~L.~Karp and F.~Mansouri,
Phys.\ Lett.\ B {\bf 480} (2000) 213
[arXiv:hep-th/0002085].


\bibitem{Mkrtchian:1982us}
R.~L.~Mkrtchian,
Nucl.\ Phys.\ B {\bf 198} (1982) 295.


\bibitem{DDG}
K.~R.~Dienes, E.~Dudas and T.~Gherghetta,
Phys.\ Rev.\ D {\bf 62}, 105023 (2000)
[arXiv:hep-ph/9912455].


\bibitem{WessBagger}
{\it Supersymmetry and Supergravity}, J. Wess and J. Bagger, Princeton
University Press (1992).


\end{thebibliography}
\end{document}